\documentclass[aps,prl,twocolumn,showpacs,superscriptaddress]{revtex4}

\usepackage{times} 
\usepackage{graphicx}
\usepackage{amsmath,amsthm,amssymb}
\usepackage{amssymb}
\usepackage{bm}
\usepackage{hyperref}
\usepackage{float}
\usepackage[usenames, dvipsnames]{color}

\begin{document} 

\title{$f(R)$-gravity after the detection of the orbital precession of the S2 star around the Galactic centre massive black hole}

\author{Ivan De Martino }
\email{ivan.demartino@usal.es}
\affiliation{Universidad de Salamanca,Facultad de Ciencias.,F\'isica Te\'orica, Salamanca, Plaza de la Merced s/n. 37008, Spain}
\affiliation{ Dipartimento di Fisica, Universit\`a di Torino,  Via P. Giuria 1, I-10125 Torino, Italy.}
\author{Riccardo della Monica}
\email{rdellamonica@usal.es}
\affiliation{Universidad de Salamanca,Facultad de Ciencias.,F\'isica Te\'orica, Salamanca, Plaza de la Merced s/n. 37008, Spain}
\author{Mariafelicia De Laurentis}
\email{mariafelicia.delaurentis@unina.it}
\affiliation{Dipartimento di Fisica, Universit\'a
di Napoli {}``Federico II'', Compl. Univ. di
Monte S. Angelo, Edificio G, Via Cinthia, I-80126, Napoli, Italy}
\affiliation{INFN Sezione  di Napoli, Compl. Univ. di
Monte S. Angelo, Edificio G, Via Cinthia, I-80126, Napoli, Italy.}
\date{}

\begin{abstract}
The GRAVITY Collaboration achieved the remarkable detection of the orbital precession of the S2 star around the Galactic centre supermassive black hole, providing yet another proof of the validity of the General Relativity. The departure from the Schwarzschild precession is encoded in the parameter $f_{\rm SP}$ which multiplies the predicted general relativistic precession. Such a parameter results to be $f_{\rm SP}=1.10\pm0.19$, which is consistent with General Relativity ($f_{\rm SP}=1$) at 1$\sigma$ level. Nevertheless, this parameter may also hide an effect of modified theories of gravity. Thus, we consider the orbital precession due to the Yukawa-like gravitational potential arising in the weak field limit of $f(R)$-gravity, and we use the current bound on the $f_{\rm SP}$  to constrain the strength and the scale length of the Yukawa-like potential. No deviation from GR are revealed at scale of  $\lambda<6300$ AU with the strength of the Yukawa potential restricted to $\delta = -0.01_{-0.14}^{+0.61}$.
\end{abstract}

\maketitle

Extensions of General Relativity (GR) are a compelling choice for providing an explanation to the ongoing accelerated expansion of the Universe, and the formation of self-gravitating systems without resorting to exotic and still unknown fluids such as Dark Matter and Dark Energy (for comprehensive reviews see \cite{Bertone2018,my_review} and \cite{Brax_2017,Huterer2018}, respectively). These extensions can be obtained replacing the Hilbert-Einstein action with a more general Lagrangian which may include  higher-order curvature invariants, such as $R^{2}$, $R_{\mu\nu} R^{\mu\nu}$, 
$R^{\mu\nu\alpha\beta}R_{\mu\nu\alpha\beta}$, $R \,\Box R$, or $R \,\Box^{k}R$, and minimally or non-minimally coupled terms between scalar fields and geometry, such as $\phi^{2}R$ \cite{defelice2010,Capozziello2011,Nojiri2011}. It is argued that extended theories of gravity must reproduce GR in their weak field limit \cite{Clifton2011}. Nevertheless, these theories represent a large collection of models that can be developed on the basis of the curvature invariants considered, and of the coupling with matter (for a comprehensive review see for instance \cite{defelice2010,Capozziello2011}). 

One appealing theory is $f(R)$-gravity, where the Hilbert-Einstein action is replaced with a more general function of the Ricci scalar \cite{Sotiriou2008}. Indeed, the cosmological constant may be naturally explained as the effect of the higher order curvature terms in the field equations, with the first attempt of this type dated back in the $1980$s \cite{Starobinsky1980}, and followed by other successful ones \cite{Li2007,WayneHU2007,Starobinsky2007,Miranda2009,Gannouji2009,Amendola2008,Seokcheon2019}. Another fascinating feature of $f(R)$-gravity is that an $R^2$-term gives rise to a Yukawa-like correction to the Newtonian gravitational potential in the weak field limit \cite{Stelle1977}. These corrections may affect the astrophysical scales of galaxies and galaxy clusters (see for instance Ref. \cite{my_review} and references therein). However, there is no statistical evidence favouring $f(R)$-gravity over  GR \cite{Lazkoz2018,Hough2019} and, additionally, GR has been collecting a huge amount of successful probes over the last decades \cite{will_2018}. It is important to remember, among others, the direct detection of gravitational waves from a binary black hole merger \cite{ligo2016}, and the subsequent direct detection of gravitational waves from a binary neutron star merger \cite{ligo2017a}.  While the first event served to show the effective existence of a fundamental pillar of GR, the second event, that was accompanied by an electromagnetic emission with a time delay of $\sim1.7$ s with respect to the merger time, was later associated to the GRB $170817$A \cite{ligo2017b}  allowing to probe the Equivalence Principle and Lorentz invariance \cite{ligo2017a}, and also to exclude several alternatives to theories of gravity \cite{Baker2017,Bettoni2017,Creminelli2017,Ezquiaga2017,Sakstein2017}.

Recently, Hees et al. (2017) \cite{Hees2017} demonstrated the effectiveness of short-period stars orbiting around the supermassive black hole to constraint the strength and the scale length of the Yukawa potential due to the fifth force.{They strongly restricted the parameter space using astrometric and spectroscopic measurement of the S2 star, obtaining an upper limit on the periastron advance of $\sim 10^{-3}$ rad/yr.}

Finally, the GRAVITY Collaboration made the pioneering detection of the orbital precession of the star S2 orbiting a compact and variable X-ray, infrared, and radio source (Sgr~A*) at the centre of the Milky Way \cite{Ghez2003,Ghez2008,Genzel2010}. Sgr~A* is supposed to be the closest supermassive black hole. It is  surrounded by a cluster of stars orbiting around, whose characteristics, such as distribution and  kinematics of stars, have been obtained through radio and infrared observations \cite{Genzel2010}. The analysis of the orbital motion of those stars served to further confirm GR. The detection of the orbital precession of the S2 star of about $\delta\phi \approx 12'$ per orbital revolution has been used to constrain the parameter $f_{\rm SP}$, which multiplies the general relativistic precession and encodes any departure from the Schwarzschild metric or from GR. Its best fit value is $f_{\rm SP} = 1.10\pm0.19$ \cite{GRAVITY2020}, where $f_{\rm SP} = 1$ leads to GR, and $f_{\rm SP} = 0$ reduces to Newtonian theory. 

Here we use the results obtained in De Laurentis, de Martino, Lazkoz (2018) \cite{DeLaurentis2018}, which computed analytically the precession of a point-like star orbiting a massive and compact object under the Yukawa-like gravitational potential arising in $f(R)$-gravity, and obtain the first constraint on the strength of the gravitational potential from the aforementioned pioneering observations, and an upper limit on the graviton mass.

\section{Background equations from {\em f(R)}-gravity}

The $f(R)$-gravity field equations are obtained by varying the action with respect to the metric. The main steps are the same as in the case of the variation of the Einstein-Hilbert action but there are also some important differences. The resulting field equations include terms containing derivatives of fourth-order in the metric and, therefore, are more complicated than GR field equations which are second-order partial differential equations, and are recovered as the special case $f (R) = R$  (for more details see  \cite{Capozziello2011} and references therein).  
Generally speaking, the $f(R)$-Lagrangian must be specified to allow practical applications and constraining the model. One way around is to require that it is an analytic Taylor expandable function. In such a case, one can straightforwardly perform the Post Newtonian limit and obtain
the solution of field equations. Considering a general spherically symmetric metric, De Laurentis, de Martino, Lazkoz (2018) obtained \cite{DeLaurentis2018}:
\begin{equation}
\label{metric}
ds^2=\left[1+\Phi(r)\right]dt^2-\left[1-\Phi(r)\right]dr^2-r^2d\Omega\,,
\end{equation}
where $d\Omega$ is the solid angle, and 
\begin{equation}\label{eq:Yukpot}
 \Phi(r) = -\frac{2G M}{
(1+\delta) r c^2}\left(1+\delta e^{-\frac{r}{\lambda}}\right)\,, 
\end{equation}
is the Yukawa-like modification of the Newtonian gravitational potential. Here $G$ is the Newton gravitational constant, $M$ is the source mass, $\delta$ is a parameter of the theory (the Newton's potential is recovered when it is turned off), and it modulates the strength of the Yukawa-like potential. Finally, $\lambda$ is a scale length which naturally arises in higher order theories of gravity  \cite{Capozziello2011}.  

Relativistic equations of motion for massive particles can be obtained from the geodesic equations for time-like geodesics of the metric in Eq. \eqref{metric}:
\begin{equation}
    \label{eq:geodeiscequations}
    \frac{d^2x^\mu}{ds^2}+\Gamma^\mu_{\nu\rho}\frac{dx^\nu}{ds}\frac{dx^\rho}{ds}=0.
\end{equation}
These equations provide differential equations for the four space-time components $\{t(s), r(s), \theta(s), \phi(s)\}$, where $s$ is an affine parameter (the proper time of the star, in our case), that can be numerically integrated once the initial conditions are specified. In order to calculate the periastron shift of a point-like particle of mass $m$ around a massive object of mass $M$,  one may assume the spherically symmetric metric in Eq. \eqref{metric}  and also that $m\ll M$ to further simplify the problem. Since the  metric is symmetric about  $\theta =\pi/2$, any geodesic that begins moving in that plane will remain there indefinitely (the plane is totally geodesic). Therefore, the coordinate system may be oriented so that the orbit of the particle lies in that plane, and fixes the
 $\theta$ coordinate to be  $\pi/2$.   
 
Using the equations of the geodesics, De Laurentis, de Martino, Lazkoz (2018) \cite{DeLaurentis2018,DeLaurentis2018b}  computed the equation of orbital precession
 \begin{eqnarray}\label{eq:deltaphi}   
\Delta\phi_{\rm{Yukawa}}&=& \frac{\Delta\phi_{GR} }{(\delta +1)}\biggl( 1+
  \frac{2 \delta  G^2 M^2 }{3a^2 c^3  \left(1-e^2\right)^2}
  \nonumber\\&& -\frac{2 \pi \delta  G^2 M^2 }{a c^4  \left(1-e^2\right) \lambda }  -
  \frac{3\delta  G M}{a c^4  \left(1-e^2\right)}
    \nonumber\\&& -\frac{ \delta G^2 M^2 }{6c^4(\delta +1) \lambda ^2}+
     \frac{\delta G M }{3 \lambda c^2 }\biggr)\,,
  \end{eqnarray}
where  $a$ is the semi-major axis, $e$ is the eccentricity, and the GR contribution to the periastron advance is:
\begin{equation}\label{eq:GRprec}
    \Delta\phi_{\rm{GR}}= \frac{6 \pi  G M}{a c^2 \left(1-e^2\right)}\,.
\end{equation}
Additionally, De Laurentis, de Martino, Lazkoz (2018) \cite{DeLaurentis2018,DeLaurentis2018b}  showed, as an example case, that in the binary system composed by the S2 star and the supermassive black hole Sgr A* differences in the orbital precession between GR and $f(R)$-gravity exist but, for a reasonable range of parameters,  do not exceed 10\%. 

\section{Data and data modelling}

The stellar cluster in the Galactic centre of the Milky Way, orbiting around a central compact object, is the most recent test bench for GR. 
There are multiple evidences that such a compact object is a supermassive black hole of mass $M\approx 4\times 10^6 M_\odot$ \cite{Ghez2008}, for instance the stellar and/or gas kinematics \cite{Gualandris2010,Genzel2010,Morris2012,Falcke2013}. This allows us to approximate the orbiting stars to massive point-like objects and to use Eq. \eqref{eq:GRprec} to predict the orbital precession of a star in GR, and Eq. \eqref{eq:deltaphi} to predict the orbital precession in the Yukawa-like potential of Eq. \eqref{eq:Yukpot}.

After an observational campaign  lasting about two decades \cite{Eckart1997}, the GRAVITY Collaboration has been able to measure the orbital precession of S2 star. They used the equation of motion at first order expansion in the post-Newtonian limit \cite{Will1972,Will2008}, and parametrized the departure from the Schwarzschild metric by introducing an {\em ad hoc} factor $f_{\rm SP}$ as follows
\begin{equation}
    \Delta\phi_{\rm{per\,orbit}} = f_{\rm SP}\times\Delta\phi_{\rm{GR}}\,.
\end{equation}
This factor may include the effect related to the spin of the black hole as well as the departure from GR. Remarkably, $f_{\rm SP}=1.10\pm0.19$ \cite{GRAVITY2020} recovering GR within $1\sigma$.

The analysis uses measurements of the positions and spectra of the star S2 collected throughout several years, and including 118 measurements obtained with the Very Large Telescope (VLT) infrared camera NACO  between 2002 and 2019.7 of the position, 75 NACO  and 54 GRAVITY measurements from 2003.3 to 2019.7  and from 2016.7 to 2019.7, respectively, of the direct S2-Sgr~A* separation with rms uncertainties of 1.7 and 0.65 mas \cite{Gravity2017}, respectively,  92 spectroscopic measurements of the 2.167 $\mu$m HI  and the 2.11 $\mu$m HeI lines between 2003.3 and 2019.45 with the spectrometer SINFONI at the VLT \cite{Gravity2019}, 2 NACO spectroscopic measurements from 2003, and 3 Keck-NIRC2 spectroscopic measurements between 2000 and 2002 \cite{Do2019}. For more details we refer to Refs. \cite{Gravity2017,Gravity2018a,Gravity2018b,Gravity2019, GRAVITY2020}. These data have been processed with a Monte Carlo Markov Chain algorithm and yield to constrain the orbital parameters (see Table E.1 in \cite{GRAVITY2020}), and the orbital precession.

Here, we use Eq. \eqref{eq:deltaphi} to predict the orbital motion and the precession of the S2 star in the Yukawa-like potential of Eq. \eqref{eq:Yukpot}, and to fit it to the data. 
More details on the data and the data analysis are given in the Supplementary Materials (SM).
First, we computed the equations of motion \eqref{eq:geodeiscequations} starting from the metric given in \eqref{metric}.  Then, initial conditions $\{t(0),r(0),\theta(0),\phi(0)\}$ and numerical values of the parameters $\{\delta,\lambda\}$ are set to integrate numerically the aforementioned equations. Since GRAVITY data are not publicly available, we use the dataset published in \cite{Gillessen2017}. This dataset includes 145 astrometic measurements \cite{Hoffman1993, Lenzen1998, Rousset1998} of the position of S2 relative to the ‘Galactic Centre (GC) infrared reference system’ \cite{Plewa2015} and 44 spectroscopic measurements \cite{Ghez2003, Eisenhauer2003, Bonnet2004}, which provide with radial velocity estimates for S2 in the local standard of rest (LSR).

Predicted positions and velocities of S2 must be corrected for some effects before being compared with the data. Here, we correct our predicted orbits for the Rømer delay, and the frequency shift due to relativistic Doppler effect. Other relativistic effects could potentially modify the astrometric positions of the observed star, like the Shapiro delay, the Lense-Thirring effect on both the orbit and the photon (in the case of a rotating BH) or the gravitational lensing of the light rays emitted by the star. Nevertheless, they are not detectable with the present sensitivity \cite{grould}.

Finally, the orbit is fully determined once the parameters $
    \bigl(M_\bullet,\allowbreak R_\bullet,\allowbreak T,\allowbreak t_p,\allowbreak a,\allowbreak e,\allowbreak i,\allowbreak \Omega,\allowbreak \omega,\allowbreak  x_0,\allowbreak v_{x,0},\allowbreak y_0,\allowbreak v_{y,0},\allowbreak v_{\rm LSR},\allowbreak \delta,\allowbreak \lambda\bigr)$
have been assigned. The first two parameters, $M_\bullet$ and $R_\bullet$, describe the mass and the distance from Earth of the source of the gravitational potential in which the star moves. The seven keplerian elements provide with initial conditions for the numerical integration of the geodesic equations and with the Thiele-Innes elements necessary to project the resulting orbit in the observer's reference frame. Five additional parameters, $(x_0,v_{x,0},y_0,v_{y,0},v_{\rm LSR})$, take into account the zero-point offset and drift of the reference frame with respect to the mass centroid and the parameters $(\delta,\lambda)$ select a particular metric \eqref{metric} for $f(R)$-gravity. This results in a total of 16 parameters whose priors are given in Table I of SM, and whose posterior distributions are  sampled with a Markov Chain Monte Carlo (MCMC) algorithm. We estimated our log-likelihood as
\begin{equation}
	\log\mathcal{L} = \log\mathcal{L}_{\rm Pos} + \log\mathcal{L}_{\rm VR} + \log\mathcal{L}_{\rm Pre}
\end{equation}
where $\log\mathcal{L}_{\rm Pos}$ is the likelihood of the positional data,
\begin{equation}
	\log\mathcal{L}_{\rm Pos} = -\sum_i\frac{(x_{\rm obs}^i-x_{\rm orb}^i)^2}{{2(\kappa\sigma_{x,{\rm obs}}^i})^2}-\frac{(y_{\rm obs}^i-y_{\rm  orb}^i)^2}{{2(\kappa\sigma_{y,{\rm obs}}^i})^2},
\end{equation}
$\log\mathcal{L}_{\rm VR}$ is the likelihood of the radial velocities,
\begin{equation}
	\log\mathcal{L}_{\rm VR} = -\sum_i\frac{(\textrm{VR}_{\rm obs}^i-\textrm{VR}_{\rm orb}^i)^2}{{2(\kappa\sigma_{\textrm{VR},{\rm obs}}^i})^2}
\end{equation}
and $\log\mathcal{L}_{\rm Pre}$ is the log-likelihood of the orbital precession given by
\begin{equation}
	\log\mathcal{L}_{\rm Pre} = -\frac{(f_{\rm SP, obs}-f_{\rm SP, th})^2}{2(\kappa\sigma_{f_{\rm SP, obs}})^2}
\end{equation}
where $f_{\rm SP, th}\equiv\Delta\phi_{\rm Yukawa}/\Delta\phi_{\rm GR}$ {and $\kappa$ is an auxiliary parameter that either takes the value $\kappa = 1$, when $\mathcal{L}_{\rm Pre}$ is set to 0 (i.e. when the precession is not taken into account in our analysis), or $\kappa = \sqrt{2}$, otherwise. This is done in order not to double-count astrometric and radial velocity data points that we implicitly assume appearing twice in the log-likelihood when considering the measurement of the orbital precession (that has been done using the same dataset as we did)}.

\section{Results and discussions}
\begin{figure}
    \centering
    \includegraphics[width=\columnwidth]{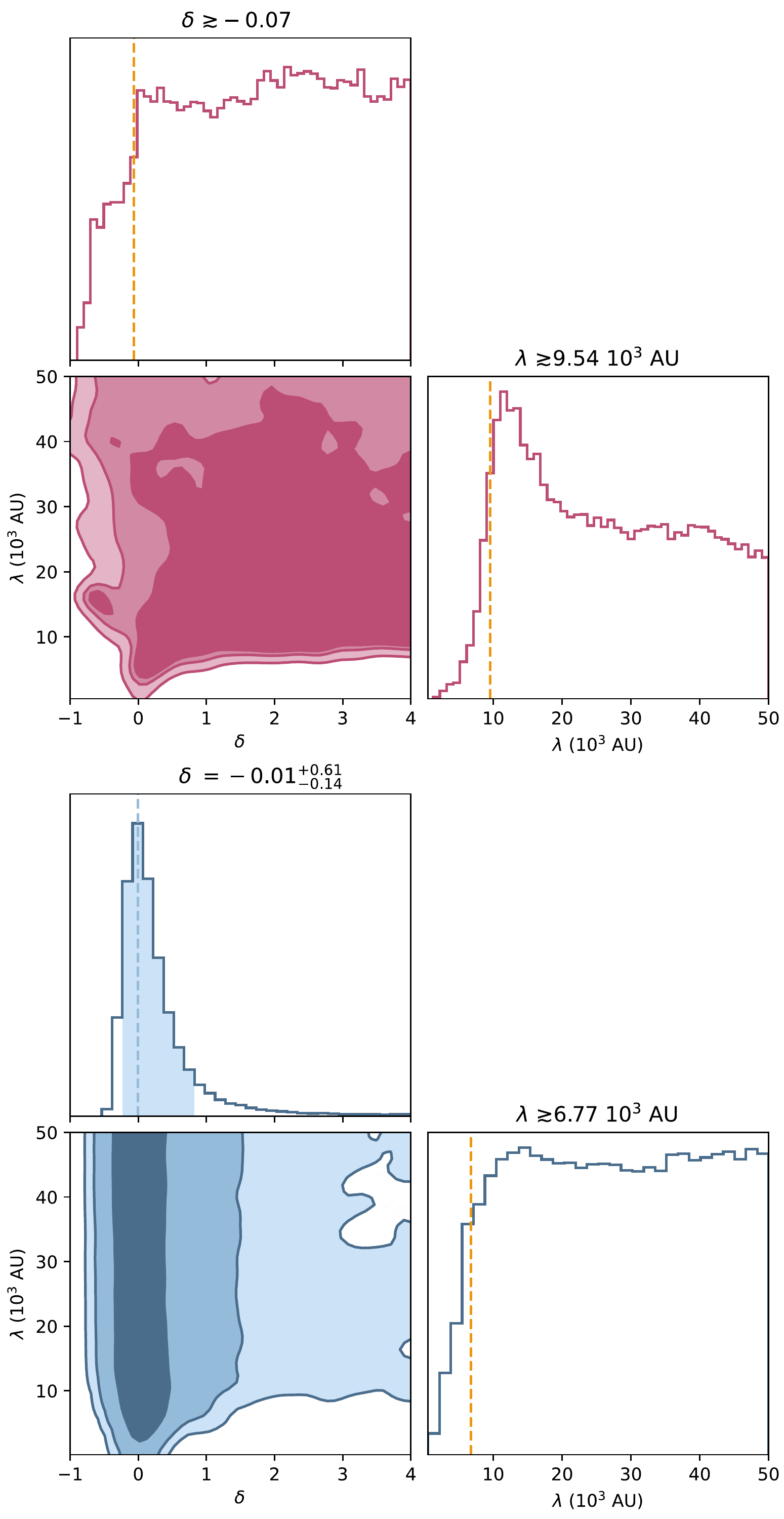}
    \caption{68\%, 95\% and 99\% confidence levels of the posterior probability density distributions for the two $f(R)$-gravity parameters $\{\delta,\lambda\}$. {\em Upper panel}: the posterior distributions are based only on data from \cite{{Gillessen2017}}. {\em Lower panel}: data includes also the measurement of the precession provided by GRAVITY Collaboration \cite{GRAVITY2020}. This is an inset of the whole corner plot presented in Figure 5 in SM.}
    \label{fig:fig1}
\end{figure}

We carried out two MCMC analysis. In the first run, we only used orbital positions and velocities, while we introduced the precession measurement in a second run. Here, we focus on the impact of our results on $f(R)$-gravity (we remand to SM for the full details). 
Figure \ref{fig:fig1} depicts the 68\%, 95\%, and 99\% confidence intervals.
Upper and lower panels illustrate  the results obtained excluding and including the measurement of the orbital precession, i.e. results in the upper panel are obtained using only measurements of position and velocities given in \cite{{Gillessen2017}}, while in the lower panels results were obtained including the measurement of the orbital precession by GRAVITY Collaboration \cite{GRAVITY2020}. The constraints on the $f(R)$-gravity parameters $\{\delta, \lambda\}$ are shown in Table I

\begin{table}[]
\setlength{\tabcolsep}{12pt}
\renewcommand{\arraystretch}{1.3}
\begin{tabular}{ccc}
\hline\hline
Parameter               & Fit w/o prec.             & Fit with prec.             \\ \hline
\bm{$\delta$}       & \bm{$\gtrsim -0,07$}    & \bm{$-0.01_{-0.14}^{+0.61}$} \\
{\bm{$\lambda$} (AU)} & \bm{$\lambda \gtrsim 9540$} & \bm{$\gtrsim 6300$}          \\ \hline\hline
\end{tabular}
\caption{Best fit values for the $f(R)$-gravity parameters using only positional and radial velocity data from \cite{{Gillessen2017}} (column 2) and using the additional measurement of the orbital precession from \cite{GRAVITY2020} (column 3).}
\label{tab:results}
\end{table}

{The additional information about the orbital precession of the S2 star provided with much tighter constraints on both $\delta$ and $\lambda$, resulting in a narrower confidence region on the $(\delta,\lambda)$ plane (see Figure \ref{fig:fig1}).  Indeed, while the analysis without the precession was not able to place an upper limit on neither $\delta$ nor $\lambda$, in the latter analysis we were able to fully constrain the parameter $\delta$, taking advantage of the greater constraining power of the orbital precession. Our analysis, thus, provides the first constraint on the strength of the Yukawa-like potential at the Galactic Centre: $\delta = -0.01_{-0.14}^{+0.61}$.}  
While, looking at the two dimensional contours, we only obtain a lower bound on the scale length of the Yukawa-like potential: $\lambda\gtrsim  6300$ AU at $1\sigma$. 
This is rather expected because this parameter is better constrained on larger astrophysical scale \cite{my_review}, and further confirms the results in \cite{DeLaurentis2018,DeLaurentis2018b}. 
{The 95\% confidence contours from our analysis are fully consistent with the exclusion regions on the fifth force determined by Hees et al. (2017) \cite{Hees2017} in the region of the parameter space that we have analysed ($\lambda > 100$ AU).}

{ We compared our results with existing constraints on $f(R)$-gravity coming from analysis at both astrophysical and cosmological scales. Specifically:
\begin{itemize}
    \item many constraints are available  at scales of the Solar System for several different $f(R)$-Lagrangians, e.g. \cite{capozziello2008} and \cite{capozziello2015}, but they are not directly comparable with our results. On the other hand, in \cite{Borka_2013}, the orbit of the S2 star is used to constrain the Yukawa-like potential, and strongly positive values of $\delta$ are favoured. It is not clear whether in \cite{Borka_2013} all {observational and relativistic effects} were taken into account. Moreover, orbits were predicted using the Newton's Law instead of integrating the geodesic equations. On the contrary we do not detect any deviation from GR, we take into account all {observational and relativistic effects} and we compute the orbits using the geodesic equations. Additionally, we converted our results in the constraints on first and second derivatives of the $f(R)$ Lagrangian. Thus, we obtain $f'(R)=0.98^{+0.26}_{-0.13}$. On the other hand, the upper limit on the scale length represents improvements of a factor $\sim 100$ with respect to  similar analysis in \cite{2014PhRvD..90d4052C}.
    \item  Our results contrast with the ones obtained for elliptical galaxies \cite{Napolitano_2012} that pointed out a severe departure from GR constraining $\delta\sim -0.8$ and $\lambda\geq10$ kpc. Also, their errors are at level of $10\%$ making it impossible to reconcile with GR, while we do not detect any departure from GR.
    \item  Our results reach the same precision of results obtained using cluster of galaxies \cite{De_Martino_2016} but contrast with \cite{capozziello2009} where authors find a value of $f'_0$ not compatible with unity and a values of $f''_0$ weakly compatible with zero (which would mean GR).  On the other hand, it is more difficult to directly compare our results with other constraints at scale of galaxy clusters since the $f(R)$-Lagrangian is different, e.g. Hu-Sawicki model, $R^n$, among the others.
\end{itemize}
}

{It is worth to remark that there are also other constraints based on PPN-parameters and pulsar timing which are at least 5-6 order of magnitude more accurate in the parameter $\delta$ than our results \cite{liu2018}. Nevertheless, our analysis is fully complementary to the other ones and can potentially reach, near in future, the required accuracy to be competitive with PPN constraints. }


\section{Conclusions}

The study of the star cluster orbiting around Sgr~A* serves to improve our knowledge on the supermassive black hole at the centre of our Galaxy, and to probe GR. Knowledge of complete orbits helps to improve the modeling the black hole itself, which is invisible in the infrared band, measuring both spin and mass of Sgr~A*. 
Currently, the stars' motion can be modelled using Newtonian physics and Kepler's laws to a high degree of accuracy, but a more detailed analysis reveals now deviations from Newtonian motion \cite{Gravity2019,GRAVITY2020}. The ever-changing motion of S2 star provides another evidence of the  Einstein's theory. The rosette effect of a star around a supermassive black hole, known as Schwarzschild precession, has been measured for the first time  \cite{GRAVITY2020}, testing GR in a new regime where gravity is stronger than in the Solar System (and even in binary pulsar systems, which provide some of the best strong-gravity tests right now) \cite{Psaltis2016,Psaltis_2011,DeLaurentisM2018}. S2 star has been studied for decades \cite{Eckart1996,Eckart1997,Ghez2003,Ghez2008,Genzel2010,Gravity2017,Gravity2018a,Gravity2018b,Gravity2019,GRAVITY2020}, and its unusual orbit was actually one of the first compelling pieces of evidence that there is a supermassive black hole at the center of the Milky Way \cite{Eckart1997}. Since it is the closest approaching star to Sgr~A*, it plays an important role for testing gravitational theories. 

Here, we have computed the orbital precession in the Yukawa-like gravitational potential arising in $f(R)$-gravity, and given in Eq. \eqref{eq:deltaphi}. 
Following the same approach of the GRAVITY Collaboration \cite{GRAVITY2020}, we have a set of 14 parameters that fully describe the orbital motion (they are accurately described in the SM). These parameters also serve to account for {observational effects, such as the offset and drift of the reference frame and the Rømer time delay, and relativistic effects, such as the gravitational and Doppler redshift.} Then we have two additional parameters, i.e. the strength and the scale length of the gravitational potential, $\delta$ and $\lambda$ respectively, which determine a possible departure from GR.
Using the remarkable measurement of the precession of the S2 star  \cite{GRAVITY2020}, we have constrained the strength $\delta = -0.01_{-0.14}^{+0.61}$, while we have set a lower limit on the scale length $\lambda>6300$ AU, as shown in Figure \ref{fig:fig1}. The latter is fully consistent with the prediction and the sensitivity analysis made in Hees et al. (2017) \cite{Hees2017}, which means no deviations from GR are measured up to this  scale. 

Indeed, other stars and interstellar medium that populates the Galactic centre affect observations \cite{Snell2011}. Additionally, other problems  are related to the Earth's atmosphere  \cite{Hardy1998}, turbulence, and absorption or refraction effects \cite{Beckers1993}.
Nevertheless, at infrared wavelengths,  photons may pass through that dust clouds unimpeded, and the motion of individual stars may be detected \cite{Ott2003,Genzel2010}. 
The outcomes of these observations include the measurement of the mass of the Milky Way's central black hole: approximately 4 million times the mass of the Sun  \cite{Eckart1996,Ghez2008}. The next horizon, quite literally, should come from the Event Horizon Telescope  \cite{Doeleman2008,Broderick2009}, a separate effort now straining to resolve the space-time right around the Milky Way’s central black hole.  
Joint observations in different bands, i.e. radio and infrared, are the only avenue towards the detection of effects related to higher order theories of gravity. Finally, we note that the future is particularly promising with higher precision radio observatories, such as SKA \cite{Keane2014} and next-generation Event Horizon Telescope \cite{Palumbo2018}, and the next generation of telescopes like the Thirty Meter Telescope, which will offer a greatly improved statistics to allow improving our constraints.

\section*{Acknowledgements}

IDM acknowledges support from Ayuda  IJCI2018-036198-I  financiada  por  MCIN/AEI/  10.13039/501100011033  y 
según  proceda:  FSE  “El  FSE  invierte  en  tu  futuro”  o  “Financiado  por  la  Unión  Europea  
“NextGenerationEU”/PRTR.  IDM  also acknowledges  the support from the grant “The Milky Way and Dwarf Weights with SpaceScales" funded by University of Torino and Compagnia di S.Paolo (UniTO-CSP),  from Junta de Castilla y León (SA096P20), and from the Spanish Ministerio de Ciencia, Innovación y Universidades and FEDER (PGC2018-096038-B-I00). RDM acknowledges support from Consejeria de Educación de la Junta de Castilla y León and from the Fondo Social Europeo.
MDL acknowledges INFN Sez. di Napoli (Iniziative SpecificaTEONGRAV and QGSKY)
\bibliography{biblio}

\end{document}


\title{Supplementary Materials} 
\author{Ivan De Martino }
\email{ivan.demartino@usal.es}
\affiliation{Universidad de Salamanca,Facultad de Ciencias.,F\'isica Te\'orica, Salamanca, Plaza de la Merced s/n. 37008, Spain}
\affiliation{ Dipartimento di Fisica, Universit\`a di Torino,  Via P. Giuria 1, I-10125 Torino, Italy.}
\author{Riccardo della Monica}
\email{rdellamonica@usal.es}
\affiliation{Universidad de Salamanca,Facultad de Ciencias.,F\'isica Te\'orica, Salamanca, Plaza de la Merced s/n. 37008, Spain}
\author{Mariafelicia De Laurentis}
\email{mariafelicia.delaurentis@unina.it}
\affiliation{Dipartimento di Fisica, Universit\'a
di Napoli {}``Federico II'', Compl. Univ. di
Monte S. Angelo, Edificio G, Via Cinthia, I-80126, Napoli, Italy}
\affiliation{INFN Sezione  di Napoli, Compl. Univ. di
Monte S. Angelo, Edificio G, Via Cinthia, I-80126, Napoli, Italy.}

\date{}

\maketitle


These Supplementary Materials have the aim of providing all the information on the analysis presented in the main document. Here, we illustrate how our pipeline works\footnote{Our pipeline is publicly available at \url{https://github.com/rdellamonica/Yukawa-S2}.}. We start describing how the numerical integration of the geodesics is performed and, then, we move to illustrate how we report the results of the integration to observable quantities. Thus, we describe both physical effects on the astrometric positions, and on the radial velocity. Finally, we accurately describe the dataset and the statistical analysis employed to constrain the physical parameters of interest.

\section{Numerical integration of mock orbits}
\label{sec:theory}
We analyse the orbital motion of the S2 star around the galactic centre in $f(R)$-gravity. To do so, we compute mock orbits using the geodesic equations in the spherically symmetric metric in \cite{delaurentisdemartino}:
\begin{equation}
    \label{eq:lineelement}
    ds^2=[1+\Phi(r)]dt^2-[1-\Phi(r)]dr^2-r^2d\Omega^2\,,
\end{equation}
where,
\begin{equation}
    \label{eq:gravpot}
    \Phi(r)=-\frac{2GM(\delta e^{-\frac{r}{\lambda}}+1)}{rc^2(\delta+1)}\,,
\end{equation}
here $G$ is Newton's gravitational constant, $M$ is the mass of the source of the gravitational field and $\delta$ and $\lambda$ are two parameters representing the strength and the scale length of the Yukawa-like modification of the gravitational potential. Relativistic equations of motion for massive particles can be obtained from the geodesic equations for time-like geodesics of the metric in Eq. \eqref{eq:lineelement}:
\begin{equation}
    \label{eq:geodeiscequations}
    \frac{d^2x^\mu}{ds^2}+\Gamma^\mu_{\nu\rho}\frac{dx^\nu}{ds}\frac{dx^\rho}{ds}=0.
\end{equation}
These equations provide differential equations for the four space-time components $\{t(s), r(s), \theta(s), \phi(s)\}$, where $s$ is an affine parameter (the proper time of the star, in our case), that can be numerically integrated once the initial conditions are specified.

\begin{figure}[!ht]
    \centering
    \begin{minipage}{.5\textwidth}
        \centering
        \includegraphics[width=.8\linewidth]{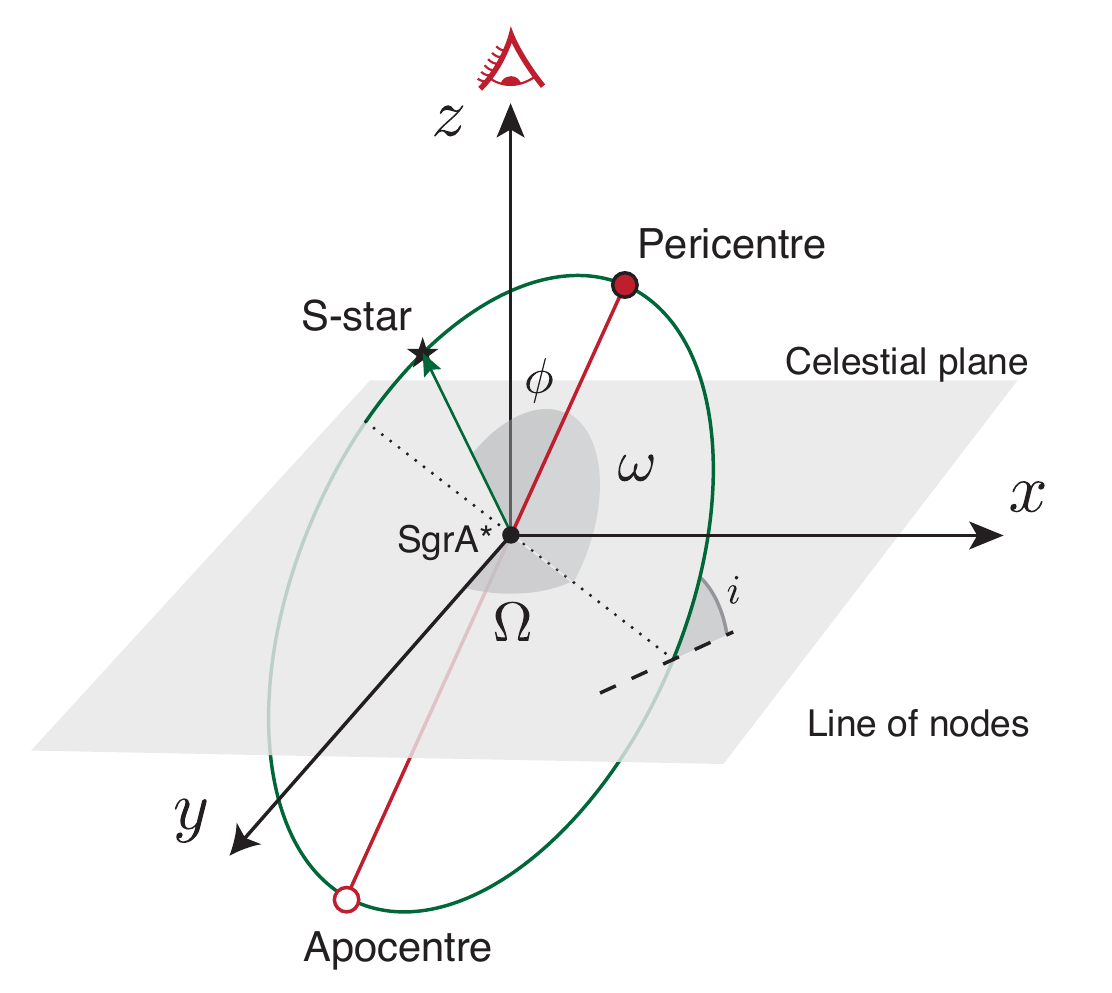}
    \end{minipage}%
    \begin{minipage}{.5\textwidth}
        \centering
        \includegraphics[width=.7\linewidth]{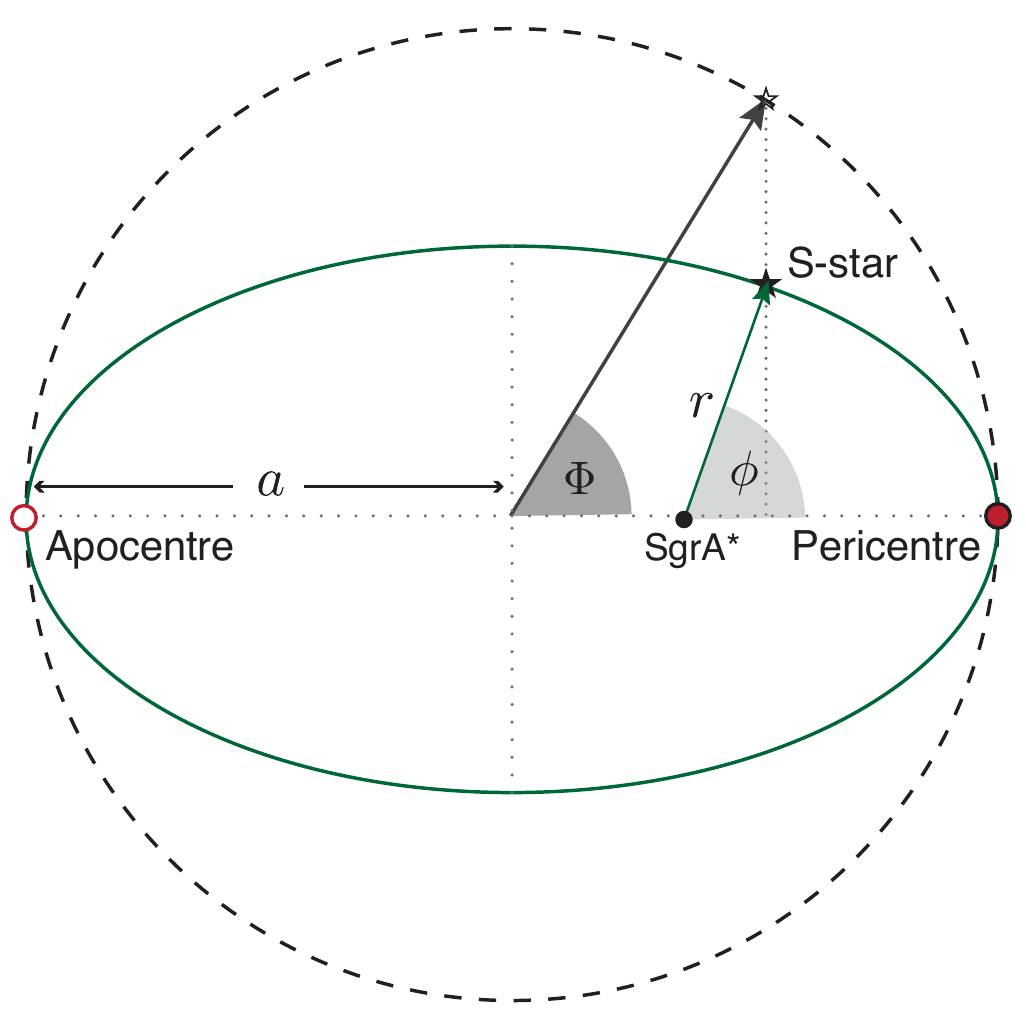}
    \end{minipage}
    \caption{Illustration of the classical keplerian elements identifying the position of a star on its orbit that is tilted with respect to an observer's reference frame (whose line of sight is aligned with the $z$ axis).}
    \label{fig:orbital_elements}
\end{figure}

First, a symbolic routine provides with the equations of motion \eqref{eq:geodeiscequations} starting from the metric given in \eqref{eq:lineelement}. At this stage $\delta$ and $\lambda$ are symbolic elements, whose values are either not fixed nor varied. 
Then, initial conditions $\{t(0),r(0),\theta(0),\phi(0)\}$ and numerical values of the parameters are set to integrate numerically the aforementioned equations. The algorithm employs an explicit Runge-Kutta 5(4) method with an adaptive step-size. 
A convenient way to place the initial conditions is to assume that $\theta(0) = \pi/2$ (the star initially lies on the equatorial plane of the reference system) and that $\dot{\theta}=0$ (its initial velocity is initially parallel to the equatorial plane). Using these hypothesis,  $\ddot{\theta} = 0$ identically \cite{delaurentisdemartino}, hence the particle moves entirely on the equatorial plane. 
Initial conditions for $r$ and $\phi$, on the other hand, can be inferred from the orbital elements \cite{green1985spherical} describing the Keplerian orbit of the star: the semi-major axis $a$, the eccentricity $e$, the inclination $i$, the angle of the line of node $\Omega$, the angle from ascending node to pericentre $\omega$, the orbital period $T$ and the time of the pericentre passage $t_p$, as illustrated in the Figure \ref{fig:orbital_elements}. 
Since a precession of the pericentre is expected, these orbital parameters are time-dependent. Therefore, they should be interpreted as the ones of the osculating ellipse at the initial time. 
In particular, we set for simplicity the initial conditions at the time of passage at apocentre $t_a = t_p-T/2$, where $\dot{r} = 0$. Therefore, both the true anomaly $\phi$ and the eccentric anomaly $\Phi$ are equal to $\pi$. Thus, the Cartesian coordinates of the star, expressed in the orbital plane, at the initial time are:
\begin{align}
    (x_{\rm orb}, y_{\rm orb}) &= \left(a(\cos\Phi-e),\, a\sqrt{1-e^2}\sin\Phi\right)=(-a(1+e),0),
\end{align}
while the Cartesian components of the velocity in the equatorial plane are:
\begin{align}
    (v_{\rm x, orb}, v_{\rm y,orb}) &= \left(-\frac{2\pi}{T}\frac{a^2}{r}\sin\Phi,\, \frac{2\pi}{T}\frac{a^2}{r}\sqrt{1-e^2}\cos\Phi\right)=
    \left(0,\, \frac{2\pi}{T}\frac{a^2}{r}\sqrt{1-e^2}\right).
\end{align}
Using these quantities, we retrieve the initial radial and angular coordinates on the orbital plane (Figure \ref{fig:orbital_elements}) and the respective velocities.
Finally, the initial condition for $\dot{t}$ descends from the normalization after requiring  $g_{\mu\nu}u^\mu u^\nu = -c^2$.

Setting the initial conditions for the passage at apocentre (which for S2 occurred on $\sim2010.35$) requires to integrate both forward and backward in time, to produce a full orbit spanning the period for which observations are available (approximately from 1992 to 2020). The integration results in two  four dimensional parametric arrays: the first one, $(t(\tau), r(\tau), \theta(\tau), \phi(\tau))$, describing the positions of S2 in the black hole's reference frame as a function of the proper time of the star; and $(\dot{t}(\tau), \dot{r}(\tau), \dot{\theta}(\tau), \dot{\phi}(\tau))$ describing its velocity.

To compare the mock orbits with the data, a projection in the reference frame of a distant observer is needed. Such a projection can be performed by means of the Thiele-Innes relations \cite{thieleinnes}: 
\begin{align}
    & A=\cos\Omega\cos\omega-\sin\Omega \sin\omega \cos i,\\
    & B=\sin\Omega \cos\omega+\cos\Omega \sin\omega \cos i\\
	& C=-\sin\omega \sin i,\\
    & F=-\cos \Omega \sin\omega-\sin\Omega \cos\omega \cos i,\\
    & G=-\sin\Omega\sin\omega+\cos\Omega \cos\omega \cos i,\\
    & H=-\cos\omega \sin i.
\end{align}
In particular, the position $(x,y)$ on the plane of the observer and the distance $z$ from it can be obtained with:
\begin{align}
     x&=Bx_{\rm orb}+Gy_{\rm orb},\\
     y&=Ax_{\rm orb}+Fy_{\rm orb},\\
     z&=Cx_{\rm orb}+Hy_{\rm orb} + R\,.
\end{align}
where $R$ is the distance of central mass from the observer and $(x_{\rm orb},y_{\rm orb})$ are the Cartesian coordinates on the orbital plane (see Figure \ref{fig:orbital_elements}), while the respective velocities read:
\begin{align}
 	v_x &= B\dot{x}_{\rm orb}+G\dot{y}_{\rm orb},\\
 	v_y &= A\dot{x}_{\rm orb}+F\dot{y}_{\rm orb},\\
	v_z &= -(C\dot{x}_{\rm orb}+H\dot{y}_{\rm orb})\,.
\end{align}
the minus sign in $v_z$ represents the fact that we interpret the radial velocity of the star as positive during the approaching phase and as negative during the recession.

\section{Connections between the predicted and the observable quantities}
A few effects (both relativistic and classical) need to be taken into account to compare the predicted positions of S2 in the reference frame of a distant observer, $(x, y, z)$, and their velocities in it, $(v_x, v_y, v_z)$,  with the  observational data.


\subsection{The Rømer delay}

When the star is farther away from the observer, photons are delayed due to the inclination of the orbit, namely Rømer delay. 
This effect modulates the time of arrival of a photon travelling from the S2 star to the Earth. As shown in \cite{gravity2018}, a first-order approximation (which is enough for our purposes, since a higher order approximation would lead to corrections of order of minutes which are negligible for our purposes \cite{do}) of such effect can be determined as
\begin{equation}
	t_{\rm em} = t_{\rm obs} - \frac{z(t_{\rm em})}{c}\approx t_{\rm obs}-\frac{z(t_{\rm obs})}{c}\left(1-\frac{v_z(t_{\rm obs})}{c}\right)\,.
	\label{eq:roemerdelay}
\end{equation}
This modulation ranges from  $\sim-0.5$ days at pericentre to $\sim7.5$ days at apocentre. Knowing the observation date $t_{\rm obs}$ from the data and using equation \eqref{eq:roemerdelay}, we determine the emission time $t_{\rm em}$ at which to compute the position on the orbit. This effect affects both the astrometric positions and measurements of the radial velocity.

\subsection{Frequency shift}

The apparent radial velocity $V_R$ of the S2 star is measured by spectroscopic observations of the photon's frequency shift
\begin{equation}
	\zeta=\frac{\Delta \nu}{\nu} = \frac{\nu_{\rm em}-\nu_{\rm obs}}{\nu_{\rm obs}}= \frac{V_R}{c}\,.
\end{equation}
The main sources of this frequency shift are the special relativistic Doppler effect, $\zeta_D$ (which for S2 is expected to be non negligible due to its high velocity at pericentre, $\approx 7650$ km/s \cite{gravity2018}),  and the gravitational redshift $\zeta_{G }$, produced by the space-time curvature in the vicinity of the mass centroid (other higher order sources of redshift, such as Shapiro time delay, can be neglected; \cite{do}). In particular, assuming that (A) the observer is in a region where the space-time curvature is negligible (which is applicable to an infinitely distant observer in an asymptotically flat space-time), and (B) the relative motion between the observer and the S2 star is totally ascribable to the motion of the star around the mass centroid (i.e. proper motion of the observer is negligible), the expressions for the various contributions to the total redshift can be written as:
\begin{align}
	\zeta = \underbrace{\frac{1}{\sqrt{-g_{00}(t_{\rm em},\vec{x}_{\rm em})}}}_{\zeta_{G}}\cdot\underbrace{\frac{\sqrt{1-\frac{v^2(t_{\rm em})}{c^2}}}{1-\vec{k}\cdot\vec{v}(t_{\rm em})}}_{\zeta_{D}}-1\,,
\end{align}
where $g_{00}$ is the coefficient of $dt^2$ in Eq. \eqref{eq:lineelement} 
, $\vec{v}$ is the space velocity of the star, and $\vec{k}$ is the spatial part of the wave vector of the observed light ray (it is worth to note that the scalar product $\vec{k}\cdot\vec{v}$ coincides with the projection of $\vec{v}$ along the line of sight, which is $v_z$). The fact that all the quantities are evaluated at $t=t_{\rm em}$ naturally accounts for the Rømer delay in the spectroscopic observables. The relativistic contributions to the total redshift amounts to $\approx200$ km/s at pericentre (where this effect is maximised because of the vicinity to the black hole and the high space velocity at the pericentre passage) and appears as a "kink" in the radial velocity with respect to what would be observed in a purely Newtonian case. Two independent observational campaigns have detected this effect for the last passage at pericentre \cite{gravity2018, do}.

\subsection{Other relativistic effects}
Other relativistic effects could potentially modify the astrometric positions of the observed star, like the Shapiro delay, the Lense-Thirring effect on both the orbit and the photon (in the case of a rotating BH) or the gravitational lensing of the light rays emitted by the star (see \cite{grould} for a in-depth analysis of all such effects). At the present sensitivity of measurements, though, \cite{grould} have shown that only the Rømer delay and the periastron advance (which is naturally considered in our treatment) are detectable.

\section{Data, data analysis, and results}
\label{sec:data}

Hereby, we describe the dataset used to carry out our analysis. These data include positions and velocities of the S2 star measured over the last three decades. Then we move to illustrate the statistical analysis employed to constrain the parameters of the model. Our analysis uses a Monte Carlo Markov Chain (MCMC) algorithm to fully explore the parameter space. Finally, we illustrate the results obtained.

\subsection{Data}
\begin{figure}[!t]
	\centering
	\includegraphics[width = \textwidth]{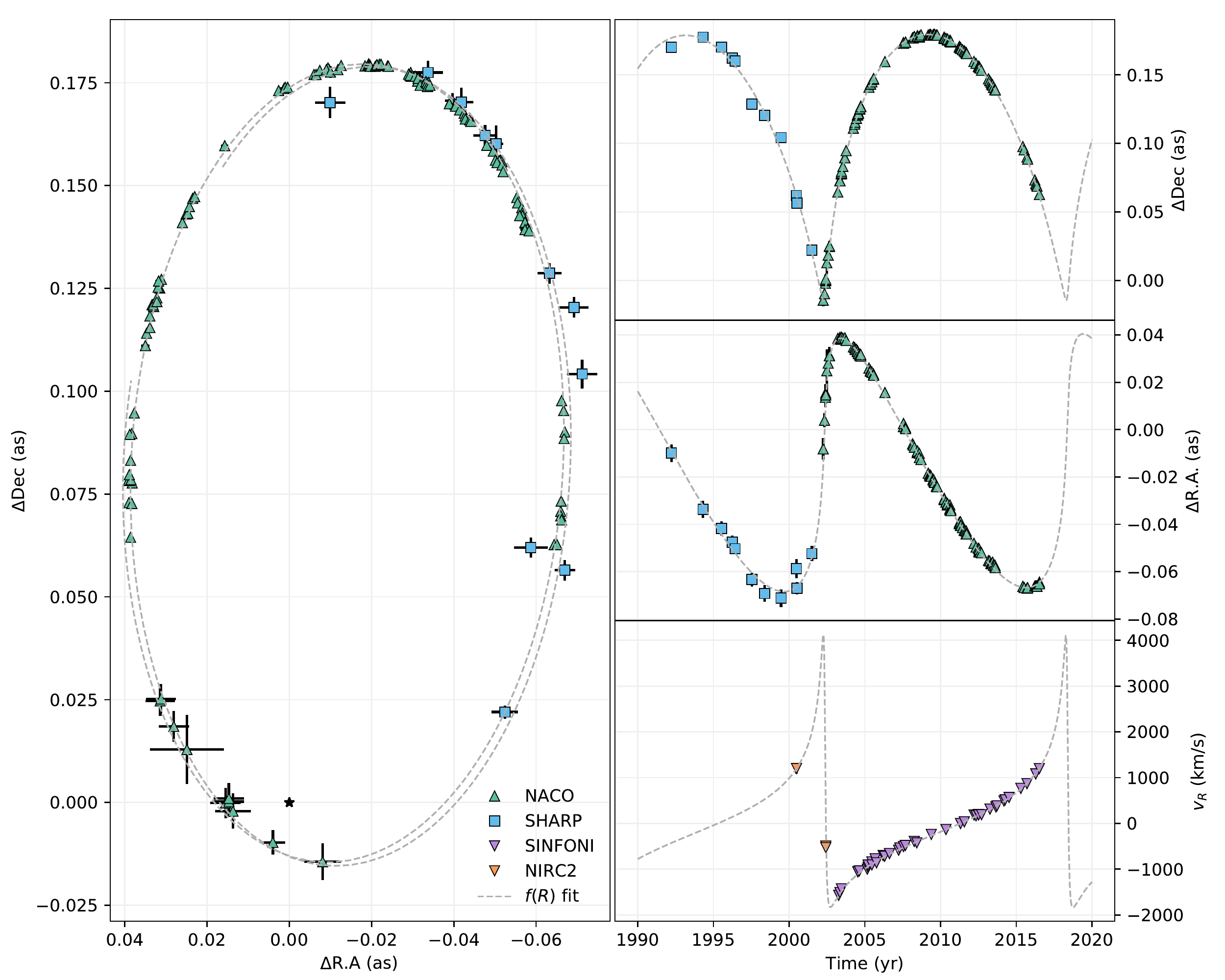}
	\caption{The dataset used for our analysis. The light blue squares and green triangles, covering the period from $\sim$1992 to $\sim2016$ are the positional data of S2 from \cite{gillessen}, whereas the violet triangles are the NIRC2 + SINFONI spectroscopic radial velocities presented in the same work. The dashed gray line is our fit from the third column in Table \ref{tab:results}. The fact that the orbit is open is ascribable to both the orbital precession and the drift of the reference frame.}
	\label{fig:datasets}
\end{figure}

Our analysis employs astrometric data that have been collected over the past three decades from different observatories. Specifically, we use the following data:
\begin{itemize}
	\item 145 data points from Table 5 in \cite{gillessen} (where an extensive description of the data analysis technique is presented) representing the position of S2 in the K or H bands, relative to the ‘Galactic Centre (GC) infrared reference system’ \cite{Plewa}, from 1992.225 to 2016.38. More precisely, data from 1992.224 to 2002 have been collected using the speckle camera SHARP at the NTT (\cite{hoffman}), which provides with astrometry whose rms uncertainty is around 3.8 mas; data from 2002 to 2016.38, on the other hand, have been collected with the VLT AO-assisted infrared camera NACO (\cite{lenzen, rousset}), whose rms uncertainty is $\approx 400 \mu$as. All data are shown in Figure \ref{fig:datasets}.
	The radio calibrations used to measure relative positions between stars in the nuclear cluster, by which the SHARP and NACO positions of S2 in the GC reference systems have been retrieved, can still be affected by a zero-point offset and a drift of the radio-reference frame with respect to the infrared reference frame. This translates into four additional parameters in the fitting procedure, $(x_0,v_{x,0},y_0,v_{y,0})$, to account for this effect. For this purpose, we use the  radio-to-infrared reference frame results from \cite{Plewa} as a prior for our MCMC, and we report them here for completeness: 
	\begin{align}
		x_0 &= -0.2\pm0.2\textrm{ mas}& 		y_0 &= 0.1\pm0.2\textrm{ mas}\nonumber\\ 
		v_{x,0} &= 0.02\pm0.1\textrm{ mas/yr}& 		v_{y,0} &= 0.06\pm0.1\textrm{ mas/yr}.
		\label{eq:drift}
	\end{align}
	\item 44 data points for the radial velocity of S2 corresponding to spectroscopic measurements of the Brackett-$\gamma$ line \cite{Gillessen_2009b} that have been collected with the AO imagers and spectrometers NIRC2 at the Keck observatory \cite{Ghez_2003}, before 2003, and with the AO-assisted integral field spectrometer SINFONI at the VLT after 2003 \cite{eisenhauer2003,bonnet}. The radial velocity measurements are corrected for the local standard of rest (LSR) \cite{gillessen} but an additional constant velocity offset, $v_{\rm LSR}$, accounting for possibile systematics is considered in our fit.
	\item Besides positional data, in \cite{gravity} a measurement of the departure from the Schwarzschild precession as the parameter $f_{\rm SP} = 1.10 \pm 0.19$ (where $f=0$ is consistent with Newtonian's gravity and $f = 1$ with GR) is found as a result of their analysis. We used this result as an additional constraint. 
\end{itemize}

\subsection{Data analysis}

The orbit of the S2 star in $f(R)$-gravity is fully determined once the following parameters are set: 
\begin{gather}
    (\underbrace{M_\bullet,R_\bullet}_{\tikzmark{a}},\underbrace{T,t_p,a,e,i,\Omega,\omega}_{\tikzmark{b}},\underbrace{x_0,v_{x,0},y_0,v_{y,0},v_{\rm LSR}}_{\tikzmark{c}},\underbrace{\delta,\lambda}_{\tikzmark{d}})\,.\\
    \nonumber
  \tikz[remember picture]{\node[align=center, anchor = north](A){\baselineskip=3pt Mass and distance \\ of central object};}\qquad
  \tikz[remember picture]{\node[align=center](B){Keplerian elements};}\qquad
  \tikz[remember picture]{\node[align=center](C){Reference frame};}\qquad
  \tikz[remember picture]{\node[align=center](D){f(R)-gravity};}\qquad
  \tikz[remember picture,overlay]{
    \draw[->] (a.south)++(0,1.5ex) to (A.north) ;
    \draw[->] (b.south)++(0,1.5ex) to (B.north) ;
    \draw[->] (c.south)++(0,1.5ex) to (C.north) ;
    \draw[->] (d.south)++(0,1.5ex) to (D.north) ;
  }  
\end{gather}  

Here, the first two parameters, $M_\bullet$ and $R_\bullet$, describe the mass and the distance from Earth of the source of the gravitational potential in which the star moves,  the seven Keplerian elements provide with initial conditions for the numerical integration of the geodesic equations, and the Thiele-Innes elements needed to project the predicted orbit in the observer's reference frame. Then, five additional parameters, $(x_0,v_{x,0},y_0,v_{y,0},v_{\rm LSR})$, account for the zero-point offset and drift of the reference frame with respect to the mass centroid, and the last two parameters, $(\delta,\lambda)$, constrain the $f(R)$-gravity model. Thus, we have 16 parameters that entirely describe the model we aim at fitting to the aforementioned data. To accomplish this objective, we explore the parameter space  using a Markov Chain Monte Carlo (MCMC) algorithm implemented in \texttt{emcee}  \cite{emcee}. 
{Even though we don't expect large deviations from GR in the values of the orbital parameters of the S2 star, we preferred to take an agnostic approach and we used uniform prior distributions on all the orbital parameters\footnote{Except for the orbital period, $T$, and the time of the pericentre passage, $t_p$, for which we adopted gaussian priors centered on the values mesaured by \cite{gravity}, due to the fact that our data do not include the astrometric position or radial velocities recorded at the pericentre passage and otherwise these parameters cannot be properly constrained. Nevertheless, in our priors we multiplied by a factor 10 the FWHM from \cite{gravity} in order not to force our results on others coming from a different model (1PN GR).} of the S2 star and for the the mass and distance of the central objects, in order not to bias the results. For the parameters related to the radio-to-infrared offset of the reference frame, on the other hand, we used gaussian priors from \cite{Plewa}. More specifically, in Table \ref{tab:priors} we provide with the complete set of priors.}
The bayesian sampler initialises 32 random walkers (i.e. 32 distinct Markov chains) 
and draws, for each of them, random samples for the 16 parameters. These are passed to our pipeline which computes the corresponding model for that particular set of parameters and finally returns to the sampler the value of the likelihood which we estimate as follows: 
\begin{equation}
	\log\mathcal{L} = \log\mathcal{L}_{\rm Pos} + \log\mathcal{L}_{\rm VR} + \log\mathcal{L}_{\rm Pre}
	\label{eq:likelihood_tot}
\end{equation}
where $\log\mathcal{L}_{\rm Pos}$ is the likelihood of the positional data,
\begin{equation}
	\log\mathcal{L}_{\rm Pos} = -\sum_i\frac{(x_{\rm obs}^i-x_{\rm orb}^i)^2}{{2(\kappa\sigma_{x,{\rm obs}}^i})^2}-\sum_i\frac{(y_{\rm obs}^i-y_{\rm  orb}^i)^2}{{2(\kappa\sigma_{y,{\rm obs}}^i})^2},
	\label{eq:likelihood_pos}
\end{equation}
$\log\mathcal{L}_{\rm VR}$ is the likelihood of the radial velocities,
\begin{equation}
	\log\mathcal{L}_{\rm VR} = -\sum_i\frac{(\textrm{VR}_{\rm obs}^i-\textrm{VR}_{\rm orb}^i)^2}{{2(\kappa\sigma_{\textrm{VR},{\rm obs}}^i})^2}
	\label{eq:likelihood_vr}
\end{equation}
and $\log\mathcal{L}_{\rm Pre}$ is the log-likelihood of the orbital precession given by
\begin{equation}
	\log\mathcal{L}_{\rm Pre} = -\frac{(f_{\rm SP, obs}-\Delta\phi_{\rm Yukawa}/\Delta\phi_{\rm GR})^2}{2(\kappa\sigma_{f_{\rm SP, obs}}^2)}
	\label{eq:likelihood_fsp}
\end{equation}
where $f_{\rm SP, obs}$ and $\sigma_{f_{\rm SP, obs}}$ are the measurements of the parameter $f_{\rm SP}$ and the corresponding statistical uncertainty from \cite{gravity}, while $\Delta\phi_{\rm Yukawa}$ and $\Delta\phi_{\rm GR}$ are the theoretically predicted values of the orbital precession in the Yukawa-like potential of $f(R)$-gravity and GR (whose expressions are given in eqs. (4) and (5) of the main paper, respectively). {The auxiliary parameter $\kappa$ either takes the value $\kappa = 1$, when $\mathcal{L}_{\rm Pre}$ is set to 0 (i.e. when the precession is not taken into account in our analysis), or $\kappa = \sqrt{2}$, otherwise. This is done in order not to double-count astrometric and radial velocity data points that we implicitly assume appearing twice in the log-likelihood when considering the measurement of the orbital precession (that has been obtained  using the same dataset as we did)}.\par
Finally, the code is run in parallel on an 8-core virtual machine that is hosted by Google Cloud Platform, which runs at an average rate of 1.50 iteration/second. Convergence is guaranteed computing the autocorrelation time, $\tau$, of the random walkers \cite{Goodman2010} and ensuring that the number of iterations of the sampler is $>F\cdot\tau$ (where in our case $F=100$) and that the estimates of $\tau$ does not vary (on average) more than 1$\%$. \cite{emcee}.
After $\approx250\,000$ iterations convergence was reached.

\begin{table}[]
\centering
\setlength{\tabcolsep}{18pt}
\renewcommand{\arraystretch}{1.5}
\begin{tabular}{lccc}
\hline
\multicolumn{1}{c}{}   &       & \multicolumn{2}{c}{Uniform priors}    \\\cline{3-4}
Parameter & & Start        & End           \\ \hline
$M_\bullet$ ($10^6M_\odot$)& & 3 & 5\\
$R_\bullet$ (kpc)& & 6.5 & 9.5\\
$a$ (mas) & &0.115 & 0.136\\
$e$ & & 0.85 & 0.9\\
$i$ $(\,^\circ)$ & & 130 & 138\\
$\Omega$ $(\,^\circ)$ & & 223 & 231\\
$\omega$ $(\,^\circ)$ & & 60 & 70\\
$v_{\rm LSR}$ (km/s) & & -50 & 50\\
$\delta$                 &     &   -0.9  & 2       \\
$\lambda$ (AU)           &     &     100 & 50$\,$000 \\ \hline
\multicolumn{1}{c}{}      &     \multicolumn{3}{c}{Gaussian priors}    \\\cline{2-4}
 & Ref. & $\sigma$        & $\mu$           \\ \hline
$T$ (yr)& \cite{gravity} & 16.0455 & 0.013 \\
$t_p - 2018$ (yr)& \cite{gravity} & 2018.37900 & 0.0016\\
$x_0$ (mas)& \cite{Plewa} & -0.2 & 0.01\\
$y_0$ (mas)& \cite{Plewa}  & 0.1 & 0.2\\
$v_{x,0}$ (mas/yr)& \cite{Plewa} & 0.05 & 0.1\\
$v_{y,0}$ (mas/yr)& \cite{Plewa}  & 0.06 & 0.1\\
\hline
\end{tabular}
\caption{The set of priors used for our analysis. {In the sub-table with the gaussian priors, the FWHM of the values from \cite{gravity} have been enlarged by a factor 10.}}
\label{tab:priors}
\end{table}

\subsection{Results and Discussions}
\begin{figure}
	\centering
	\includegraphics[width = \textwidth]{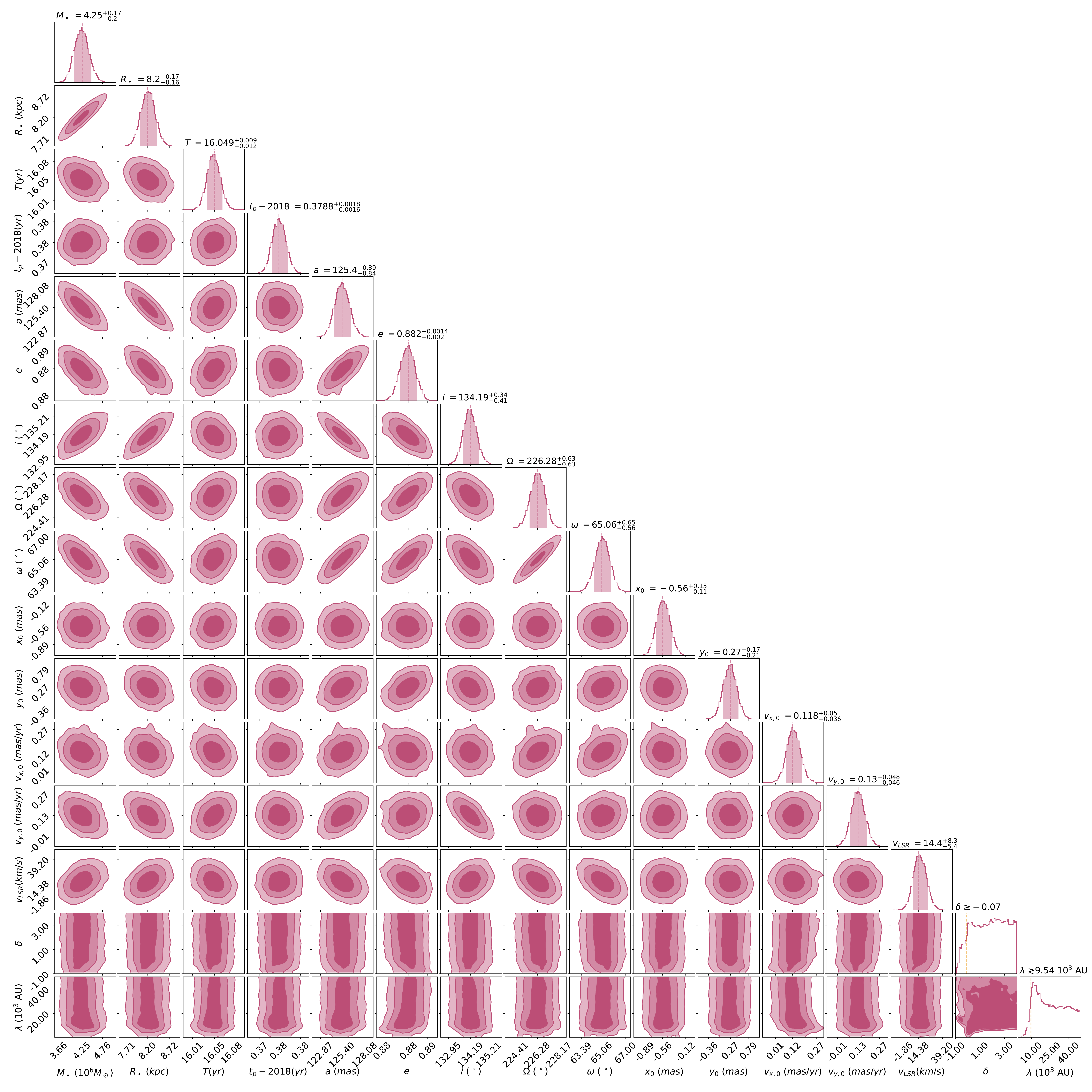}
	\caption{68\%, 95\% and 99\% confidence levels of the posterior probability density distributions for the entire set of parameters, but including only data from \cite{gillessen}.}
	\label{fig:mcmc_gillessen}
\end{figure}

\begin{figure}
	\centering
	\includegraphics[width = \textwidth]{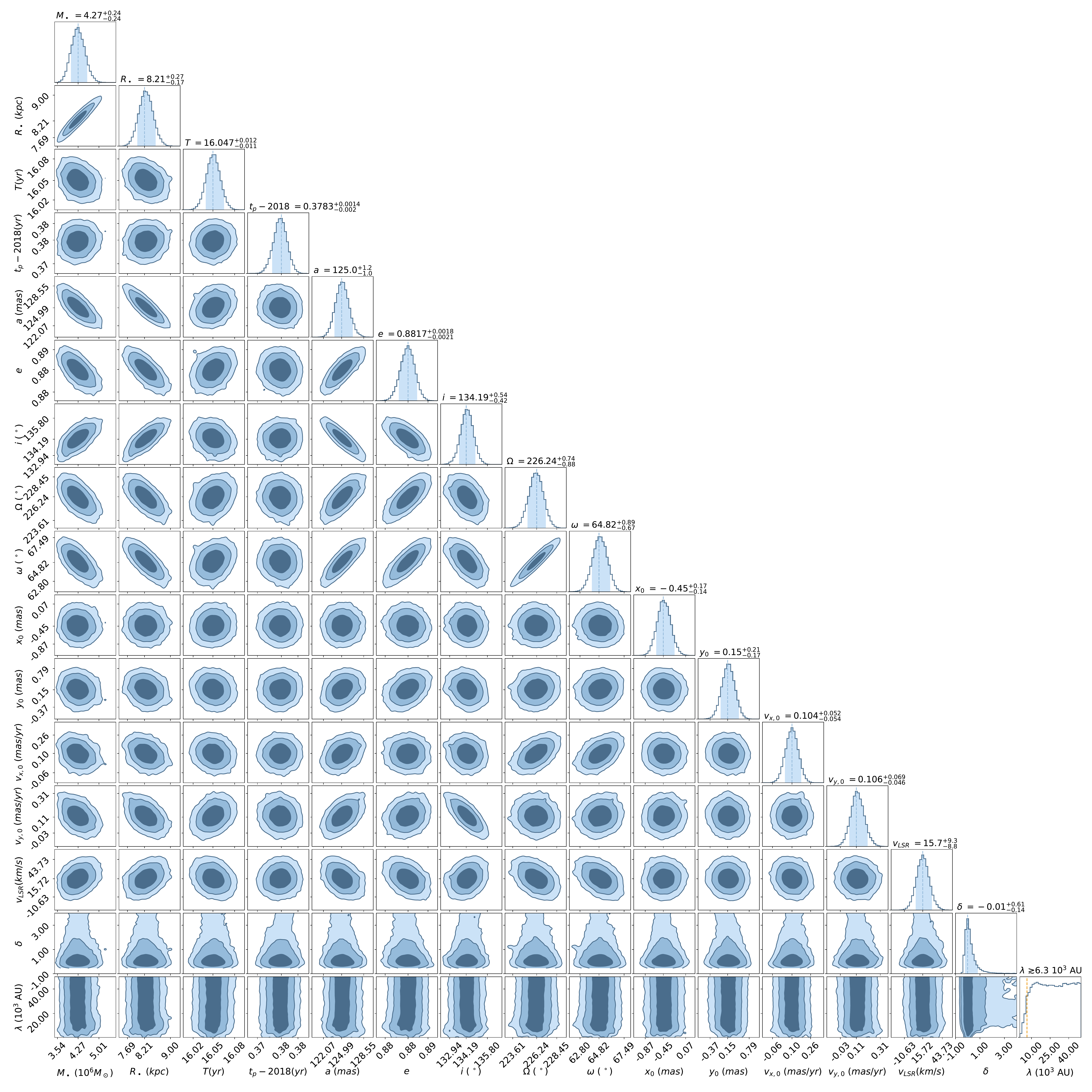}
	\caption{The same of Figure \ref{fig:mcmc_gillessen} but including also the measurement of the precession provided by GRAVITY Collaboration \cite{gravity}.}
	\label{fig:mcmc_gravity}
\end{figure}

\hypertarget{sec:results}{}
We carried out two different samplings of the posterior distributions that we hereby report:
\begin{enumerate}
	\item we run the MCMC analysis using only data from \cite{gillessen}, without using the precession information. The complete posterior distribution of the 16 parameters is then reported in Figure \ref{fig:mcmc_gillessen} and the corresponding best fit values for the parameters are reported in the second column of Table \ref{tab:results}.
	\item we run the MCMC analysis using data from \cite{gillessen} and the precession information from \cite{gravity}. The results are shown in Figure \ref{fig:mcmc_gravity} with the corresponding best-fit values reported in the third column of Table \ref{tab:results}. 
\end{enumerate}
 
\begin{figure}
    \centering
    \includegraphics[width = \textwidth]{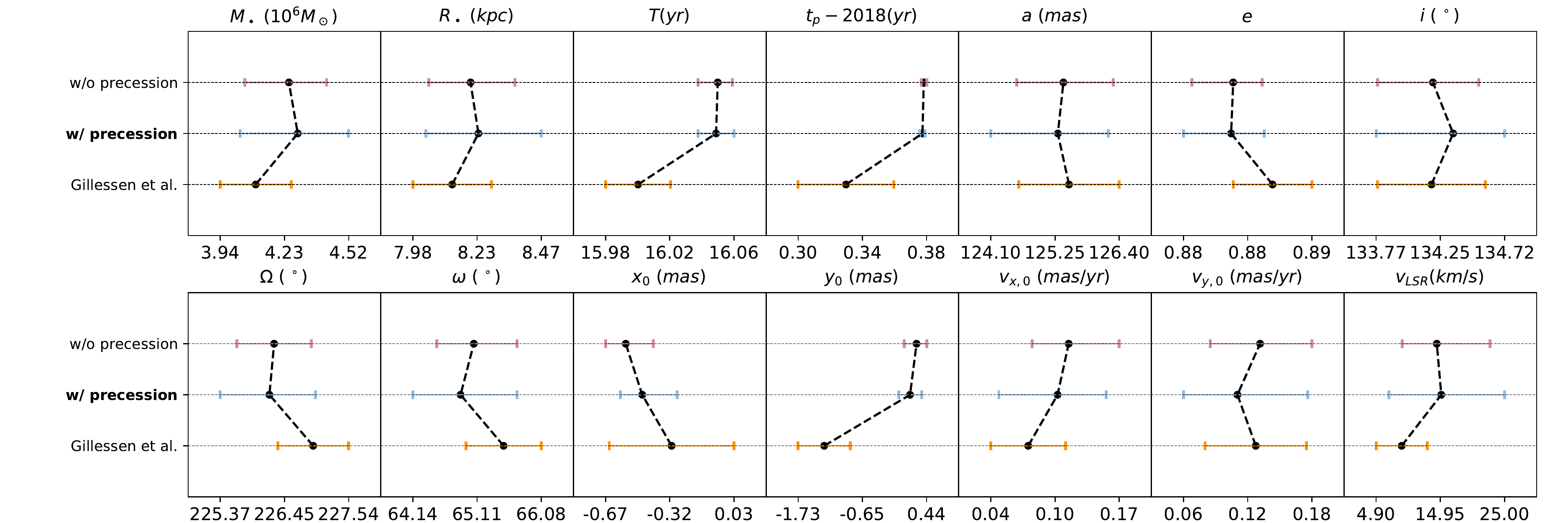}
    \caption{A comparison between our best fit orbital parameters with the corresponding 1$\sigma$ c.i. for both the case without (top row) and with (central row) the information of the precession at the pericentre passage, and the best fitting parameters from \cite{gillessen} (bottom row). Refer to Table \ref{tab:results} for the particular set of parameters from \cite{gillessen} that we choose to compare with.}
    \label{fig:comparison}
\end{figure}

{The additional information about the orbital precession of the S2 star provided with much tighter constraints on both $\delta$ and $\lambda$, resulting in a narrower confidence region on the $(\delta,\lambda)$ plane (see Figure 1 in the main draft). Indeed, while the analysis without the precession was not able to place an upper limit on neither $\delta$ nor $\lambda$ (we could only place lower limits on these parameters as reported in the last two rows of Table \ref{tab:results}), in the latter analysis we were able to fully constrain the parameter $\delta$, taking advantage of the greater constraining power of  the orbital precession. Moreover, the best fit values of Gillenssen et al. \cite{gillessen}, that we report in the fourth column of Table \ref{tab:results}, fall inside our $2\sigma$ confidence region, as shown in Figure \ref{fig:comparison} where they are reported in the third row and directly compared to our best fitting results. 
This was rather expected. Indeed, since we do not detect any departure from GR, we expect our best fits to be close to ones from standard analyses.}

\begin{table}[]
\setlength{\tabcolsep}{18pt}
\renewcommand{\arraystretch}{1.5}
\begin{tabular}{lccc}
\hline\hline
Parameter               & Fit w/o precession              & Fit with precession              & Best-fit values from \cite{gillessen} \\ \hline
$M_\bullet$ ($10^6M_\odot$) &  $4.24\pm0.18$ & $4.28_{-0.25}^{+0.24}$ & $4.1\pm0.16$  \\
$R_\bullet$ (kpc) &  $8.21\pm0.17$ & $8.26\pm0.22$ & $8.13\pm0.15$  \\
$T$ (yr) &  $16.048\pm0.01$ & $16.048\pm0.011$ & $16.0\pm0.02$  \\
$t_p - 2018$ (yr) &  $0.3789\pm0.0017$ & $0.378\pm0.0017$ & $0.33\pm0.03$  \\
$a$ (mas) &  $125.42\pm0.87$ & $125.1\pm1.1$ & $125.5\pm0.9$  \\
$e$ () &  $0.8817\pm0.0017$ & $0.8815\pm0.002$ & $0.8839\pm0.0019$  \\
$i$ ($.^\circ$) &  $134.15_{-0.37}^{+0.38}$ & $134.25\pm0.48$ & $134.18\pm0.4$  \\
$\Omega$ ($.^\circ$) &  $226.29\pm0.63$ & $226.17_{-0.8}^{+0.81}$ & $226.94\pm0.6$  \\
$\omega$ ($.^\circ$) &  $65.1\pm0.6$ & $64.92\pm0.78$ & $65.51\pm0.57$  \\
$x_0$ (mas) &  $-0.54\pm0.13$ & $-0.43\pm0.16$ & $-0.31\pm0.34$  \\
$y_0$ (mas) &  $0.25\pm0.19$ & $0.17\pm0.19$ & $-1.29\pm0.44$  \\
$v_{x,0}$ (mas/yr) &  $0.125\pm0.043$ & $0.102\pm0.053$ & $0.078\pm0.037$  \\
$v_{y,0}$ (mas/yr) &  $0.131_{-0.046}^{+0.047}$ & $0.117_{-0.057}^{+0.058}$ & $0.126\pm0.047$  \\
$v_{\rm LSR}$ (km/s) &  $15.8_{-6.8}^{+6.9}$ & $16.0_{-9.0}^{+9.1}$ & $8.9\pm4.0$  \\ \hline                          
\bm{$\delta$}       & \bm{$\gtrsim -0.07 $}    & \bm{$-0.01_{-0.14}^{+0.61}$} & \multicolumn{1}{l}{}                                   \\
\bm{$\lambda$} {(AU)} & \bm{$\lambda \gtrsim 9540$} & \bm{$\gtrsim 6300$}          & \multicolumn{1}{l}{}                                   \\ \hline\hline
\multicolumn{4}{p{0.9\textwidth}}{$^*$ The actual value of $t_p$ reported in \cite{gillessen} is 2002.33. We have shifted it by an orbital period $T$ in order to make it comparable to our estimates.} 
\end{tabular}
\caption{{\em Column 1:} the 16 free parameters used in our analysis; {\em Column 2:} best fit values obtained using only positional and radial velocity data from \cite{gillessen}; {\em Column 3:} best fits obtained with the additional information of the the orbital precession from \cite{gravity}; {\em Column 4:} the best fit values from Gillenssen et al. \cite{gillessen}. In particular, the values of $M_\bullet$ and $R_\bullet$ refer to the second row of their Table 1 (determined using only observations from VLT and with priors from \cite{Plewa} on the reference frame) and the values for the orbital parameters are reported in their Table 3 for S2.}
\label{tab:results}
\end{table}

\bibliography{biblio_SM}
